# Broadband quantum spectroscopy at the fingerprint mid-infrared region

Anna V. Paterova*, Zi S. D. Toa, Hongzhi Yang, and Leonid A. Krivitsky

*Institute of Material Research and Engineering, Agency for Science Technology and Research (A\*STAR), 138634 Singapore*

*E-mail: Paterova_Anna@imre.a-star.edu.sg

**Abstract**

Numerous molecules exhibit unique absorption bands in the "fingerprint" infrared (IR) region, allowing chemical identification and detection. This makes IR spectroscopy an essential tool in biomedical diagnostics, sensing, and material characterization. However, measurements in the IR range are challenging due to the relatively poor performance, need for cryogenic cooling, and high cost of IR light sources and photodetectors. Here, we demonstrate that the mid-IR fingerprints of the sample can be revealed from measurements in the near-IR range using conventional silicon photodetectors. Our approach relies on the quantum interference of correlated photon pairs produced in strongly frequency non-degenerate parametric down-conversion. Our technique simultaneously measures the absorption coefficient and refractive index in the broad "fingerprint" region with high accuracy. As a proof-of-concept, we perform spectroscopy of nitrous oxide gas in the 7.4-8.4 μm wavelength range, with the detection in the 865-877 nm range. This work extends the applicability of quantum interferometry to a range of broad practical appeals, such as detection of atmosphere pollutants, breath diagnostics, chemical safety, and others.

**Introduction**

Infrared (IR) spectroscopy in the so-called "fingerprint" mid-IR region of wavelengths (from 1500 $cm^{-1}$ to 600 $cm^{-1}$) is a useful tool in a variety of applications[1]. This region contains fundamental molecular vibrational bands, such as the C-O, C-C, and C-N stretches, the C-H bending vibrations, and some benzene ring signatures. Even though it is difficult to discern the IR signatures of specific functional groups in this spectral region, it is, as its name suggests, unique to every molecule. Thus, in addition to detection-based applications such as gas sensing and environment control, IR spectroscopy is essential for identification-based applications, such as protein spectroscopy, food quality control, forensics, and many others[2-4].

In conventional IR spectroscopy, such as Fourier transform IR spectroscopy (FTIR), the absorption of the IR light is detected directly[5]. The chemical composition of the sample is then unraveled from the obtained spectrogram. A common limitation of this approach is the low sensitivity and high background noise of IR detectors. To illustrate, mid-IR detectors, covering the spectral range from 0.7 μm to over 20 μm, typically uses HgCdTe (MCT). At room temperature, the detection efficiency of these detectors is around three orders of magnitude smaller than Si-based photodiodes[6,7].

Therefore, cryogenic cooling (down to 77K) is needed to achieve an acceptable signal-to-noise ratio. Despite continuous efforts in the community for advancing mid-IR detectors[8], a viable alternative relies on using nonlinear optical phenomena. In this case, the interaction of a specimen with the probing IR light can be revealed from the detection of the visible/near-IR (NIR) light, where efficient Si detectors can be used.

One of the first of such methods is up-conversion spectroscopy. The mid-IR spectrum of the specimen, probed by weak mid-IR light, is "transferred" to the NIR (or visible) range via sum-frequency generation using a pump laser. Often this scheme implicates a high-power pump laser (~10W) and a globar mid-IR light source[9]. However, the requirement of a high-power laser limits this technique from widespread benchtop adoption.

Another promising method uses the "induced coherence" phenomenon[10-12]. Here, two nonlinear crystals are placed in an interferometer and coherently pumped by a common laser. The crystals generate identical photon pairs via spontaneous parametric down-conversion (SPDC). By setting the phase-matching conditions, signal and idler photons of a pair can be generated in visible/NIR and mid-IR wavelengths, respectively. The paths of signal and idler photons, generated in the two crystals, are carefully superimposed, and the specimen is placed in the path of the idler photons. The key feature of this interferometer is that the interference pattern of signal photons, observed after the second crystal, depends on the phase and amplitude of idler photons that interact with the sample. Thus, from the interference pattern of visible/NIR photons, we can infer the specimen properties in the mid-IR range.

The "induced coherence" technique has already been demonstrated in several applications, including IR and THz spectroscopy[13-19], IR imaging[20-24], IR optical coherence tomography[25-28], and IR polarimetry[29]. However, this technique has not been realized in the mid-IR "fingerprint" range of 6-16 μm. Extending the quantum interferometry to the "fingerprint" mid-IR range is challenging because the mid-IR photons have large scattering angles in the strongly non-degenerate SPDC. Thus, a particular arrangement of the optical system is required to overlap the modes of signal and idler photons and then observe the quantum interference.

Here, we extend the "induced coherence" technique to the "fingerprint" mid-IR range from 7.32 μm to 8.87 μm, by designing a dedicated optical setup to accommodate strongly non-degenerate SPDC. We use silver thiogallate ($AgGaS_2$ or AGS) crystal as a source of correlated photon pairs. Our setup allows high sensitivity and accurate measurement of the gas media. Through careful optical engineering, we achieve remarkable robustness of the interferometer to external vibrations. This approach extends the application range of this method, which makes it appealing for applications in real-world settings related to biomedical, material characterization, and studies of climate change.

**Interferometer design**

The scheme of our experiment is shown in Fig. 1. The common-path interferometer is formed by the AGS crystal, followed by a 90°off-axis parabolic mirror (OAPM) and a flat mirror. The AGS crystal has a broad transparency range from 0.7 μm to 13 μm, which generates NIR and mid-IR SPDC photons by using the NIR pump.

The optical elements are arranged in a folded 4-*f* system, in which the crystal and the flat mirror are placed at focal distances from the center of the OAPM. The position of the flat mirror is fixed, while the crystal can be displaced by a small distance $\varepsilon/2$ from the initial position. The pump beam generates SPDC photon pairs, which are focused on the flat mirror by the OAPM. The photons (both SPDC and pump) are then reflected to the nonlinear crystal, where SPDC photons are generated again by the reflected pump beam. With the help of the OAPM, we accurately overlap spatial modes of the SPDC photons generated from the forward and backward passes of the pump beam through the crystal. Due to the indistinguishability of generated SPDC photon pairs[30], we observe an interference pattern across a broad wavelength-angular spectrum of the SPDC. The interference fringes of signal photons are then measured by the CCD camera of the spectrometer (not shown in the scheme, see Methods section). We should also note that our scheme is intrinsically stable since all the waves travel along a common path.

Even though the scheme resembles earlier demonstrations[13,14], we implement several significant improvements to its performance:

1. Previous IR spectroscopy realizations did not use focusing optics and hence were limited by the optical path of the probing IR light through the medium under study, typically tens of millimeters[13,14]. In contrast, our scheme has an order of magnitude longer interaction length of the IR light with the gas under study. This is achieved by using the OAPM, which effectively compensates for the divergence of signal and idler photons.
2. In the previous work, the frequency bandwidth was limited by the linewidth of the SPDC process, not exceeding 250 nm in the IR[13,14]. Tuning to the new wavelength required tilting the crystal, thus compromising the readout speed. In contrast, in our method, the interference pattern is observed at larger SPDC angles due to careful alignment of the spatial modes. Since in SPDC each angle corresponds to a specific wavelength, we realize a "one-shot" readout of the spectra across the bandwidth of 1.55 microns in the mid-IR, which significantly increases the spectral coverage.

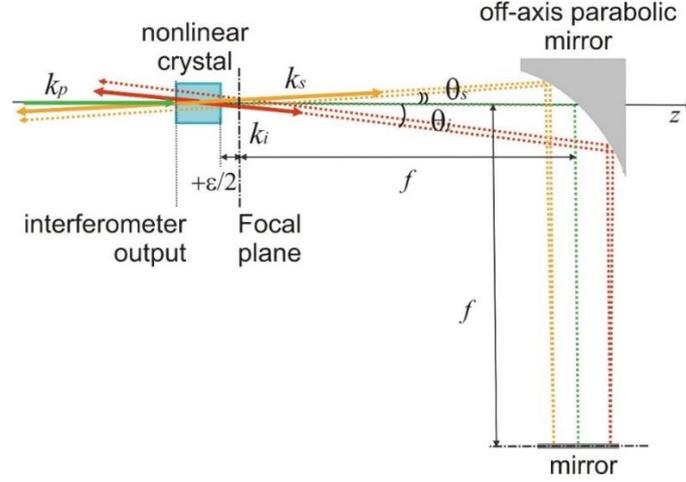

*Fig. 1 Common-path quantum interferometer with one nonlinear crystal, an off-axis parabolic (OAPM), and flat mirrors.* The crystal and the flat mirror are placed approximately at the focal distance of the OAPM mirror, where displacement $\varepsilon/2 \ll f$. The positive direction of the displacement corresponds to the increased distance between OAPM and the crystal. $\theta_{s,i}$ are scattering angles of the signal and idler photons, respectively. The z-axis is considered to be along the pump beam propagation direction. Focal planes of the OAPM are indicated by dashed-dotted lines. Dashed lines represent the optical paths of the photons inside the interferometer. When the crystal is displaced by $\varepsilon/2$ from the focal plane of OAPM, the SPDC photons generated at the forward pass of the pump through the crystal acquire a transverse phase shift with respect to the SPDC photons generated at the backward pass. The interference pattern of signal photons is detected using the spectrograph and the camera behind the crystal (not shown).

Using the proposed common-path interferometer scheme, we perform IR spectroscopy of nitrous oxide gas ($N_2O$), which has an absorption doublet at 7.69 and 7.88 μm[31]. Nitrous oxide is a good candidate here due to its non-toxicity and transparency in the visible and NIR ranges.

### Results

*Theory*

The theory of induced coherence has been studied in earlier works for the case of the time domain[10,30] and the frequency domain[32,33]. However, a theory considering the transverse overlap of modes of SPDC photons, which is relevant to our setup, has not been discussed earlier. Here we consider the interferometer scheme in Fig. 1 and describe how the inclusion of optical elements, such as lenses, OAPM, and mirror influences the interference fringes in the frequency-angular domain.

In the plane wave approximation for the pump beam and infinite nonlinear crystal in the transverse directions, the state vector of the SPDC photons is given by[34-36]:

$$|\psi\rangle \sim \eta \sum_{\mathbf{k}_s,\mathbf{k}_i} F_{\mathbf{k}_s\mathbf{k}_i} a^+_{\mathbf{k}_s} a^+_{\mathbf{k}_i} |0\rangle, \tag{1a}$$

$$F_{k_s k_i} = F_{k_s} = F_{k_i} \propto \int_{-l/2}^{l/2} dz \exp(i\Delta k z) \propto sinc(\Delta k l/2), \tag{1b}$$

where $k_j$ indicates the longitudinal components of the wave vectors for $j=p,s,i$ pump, signal and idler photons, respectively; $\eta$ is the efficiency of the SPDC generation; $a^+_{k_{s,i}}$ are photon creation operators for signal and idler photons, respectively; $|0\rangle$ is the vacuum state vector, $F_{k_s k_i}$ is the two-photon state amplitude; $\Delta k = k_p - k_s - k_i$ is the phase mismatch inside the nonlinear crystal along the $z$-axis, $l$ is the length of the nonlinear crystal.

First, let us consider single spatial modes $s$, $i$ of the SPDC generated in the forward pass of the pump beam through the crystal. In this case, the SPDC state vector after the crystal is given by:

$$|\psi_1\rangle = a^+_s a^+_i |0\rangle. \tag{2}$$

Due to the slight displacement of the crystal $\varepsilon/2$ ($\varepsilon \in \mathbb{R}$), see Fig. 1, the state vector $\psi_1$ acquires an additional phase while traveling to the focal plane of the OAPM:

$$|\psi'_1\rangle = \exp(i\varphi_s + i\varphi_i) a^+_s a^+_i |0\rangle, \tag{3}$$

where $\varphi_j = k'_j \varepsilon/2$, $k'_j$ are the wave vectors in the medium outside of the crystal, with the refractive indices $n'_j$.

The state vector at the focal plane after the OAPM (which coincides with the surface of the flat mirror), is given by the Fourier transformation $\mathcal{F}$ of the state vector in Eq. (3):

$$|\tilde{\psi}_1\rangle = \mathcal{F}[|\psi'_1\rangle] = \mathcal{F}[\exp(i\varphi_s + i\varphi_i + i\pi/2) a^+_s a^+_i |0\rangle]. \tag{4}$$

where $\pi/2$ is an additional phase shift acquired due to the rotation of the Cartesian coordinate system after reflection from the OAPM. Since the flat mirror is located at the focal distance from the OAPM, the state vector at the focal plane after the second reflection from the OAPM will be given by the inverse Fourier transformation $\mathcal{F}^{-1}$:

$$|\psi''_1\rangle = \mathcal{F}^{-1}[|\tilde{\psi}_1\rangle] = \mathcal{F}^{-1}[\mathcal{F}[\exp(i\varphi_s + i\varphi_i + i\pi/2) a^+_s a^+_i |0\rangle] \exp(i3\pi/2)] =$$

$$\exp(i\varphi_s + i\varphi_i) a^+_s a^+_i |0\rangle. \tag{5}$$

where $3\pi/2$ accounts for the phase shift upon reflection from the flat mirror and the OAPM. At the output of the interferometer, the state vector acquires an additional phase traveling $\varepsilon/2$ distance from the focal plane and through the nonlinear crystal with the length $l$:

$$|\tilde{\psi}''_1\rangle = \exp(i2\varphi_s + i2\varphi_i + ik_s l + ik_i l) a^+_s a^+_i |0\rangle. \tag{6}$$

The state vector described by Eq. (6) is similar to the state vector $\psi_1$ with an additional phase component. Next, at the backward pass of the pump through the crystal, the second pair of photons is generated. It is described by the state vector:

$$|\psi_2\rangle = \exp(-i\tilde{\varphi}_p)\, a_s^+(\tau_i^* a_i^+ + r_i^* a_{i0}^+)|0\rangle, \qquad (7)$$

where $\tilde{\varphi}_p = k'_p \varepsilon + k_p l$ is the phase acquired by the pump beam traveling inside the interferometer. Equation (7) accounts for losses in the interferometer in the idler mode using a beamsplitter model[37], with the amplitude transmission and reflection coefficients $\tau_i$ and $r_i$, respectively. Here $i_0$ denotes the state from the unused port of the beamsplitter. Then, the final state of the SPDC photons after the interferometer is given by:

$$|\Psi\rangle = |\tilde{\psi}''_1\rangle + |\psi_2\rangle = \exp(ik'_s \varepsilon + ik'_i \varepsilon + ik_s l + ik_i l) a_s^+ a_i^+ |0\rangle +$$
$$\exp(-ik'_p \varepsilon - ik_p l)\, a_s^+(\tau^* a_i^+ + r^* a_{i0}^+)|0\rangle. \qquad (8)$$

Considering Eq. (1), the state vector of the SPDC photons for all spatial modes is given by:

$$|\Psi\rangle \sim \eta \sum_{\boldsymbol{k}_s,\boldsymbol{k}_i} F_{k_s k_i} \exp(ik'_{\boldsymbol{k}_s}\varepsilon + ik'_{\boldsymbol{k}_i}\varepsilon + ik_{\boldsymbol{k}_s}l + ik_{\boldsymbol{k}_i}l) a_{\boldsymbol{k}_s}^+ a_{\boldsymbol{k}_i}^+|0\rangle +$$
$$\eta \sum_{\boldsymbol{k}_s,\boldsymbol{k}_i} F_{k_s k_i} \exp(-ik'_p \varepsilon - ik_p l)\, a_{\boldsymbol{k}_s}^+(\tau^* a_{\boldsymbol{k}_i}^+ + r^* a_{i0}^+)|0\rangle, \qquad (9)$$

In our experiment, we detect the wavelength-angular spectrum (intensity) of signal photons, which can be obtained from Eq. (9) as follows[32,38]:

$$I_s = \langle\Psi|a_{\boldsymbol{k}_s}^+ a_{\boldsymbol{k}_s}|\Psi\rangle = sinc^2(\Delta k l/2)(1 + |\tau_i|\cos(\Delta k l + \Delta k' \varepsilon)). \qquad (10)$$

where $\Delta k' = k'_p - k'_s - k'_i$ is the phase mismatch in a sample under test.

The interference pattern in Eq. (10) includes refractive index $n_i$ and amplitude transmissivity $\tau_i$ of a medium at the wavelength of idler photons. The phase shift of the pattern relative to some reference is given by the argument under the cosine function, while the visibility $V$ of the interference pattern is given by the following:

$$V = \frac{I_s^{max} - I_s^{min}}{I_s^{max} + I_s^{min}} = |\tau_i| = \exp\left(-\frac{1}{2}\alpha_i(4f + \varepsilon)\right), \qquad (11)$$

where $\alpha_i$ is the absorption coefficient of the medium at the idler photons wavelength. Thus, the refractive index and the absorption coefficient can be inferred simultaneously from phase shifts of the interference fringes and the change in interference visibility, respectively.

The visibility of the interference pattern also depends on the indistinguishability of photons pairs created in two passes of the pump beam[30,39] in the interferometer. The indistinguishability criterion in our case is determined by the displacement of the nonlinear crystal:

$$(2l + \varepsilon/2) \tan \theta_s \ll d, \tag{12}$$

where $d$ is the pump beam diameter, $\theta_s$ is the scattering angle of signal photons. As the displacement $\varepsilon$ in Eq. (12) is small, the indistinguishability criterion can be met for large scattering angles and with a much larger optical path length inside the interferometer $l'=4f+\varepsilon$. This is in striking contrast with previous works[13,14] in which the indistinguishability condition is defined by the distance between the crystals, which is limited to a few centimeters.

## *Observation of the quantum interference in the mid-IR*

In our experiments, the interferometer is placed inside a vacuum chamber. First, we align the phase-matching angle of the AGS crystal at 53.70°±0.02° and take reference measurements with the vacuum. The observed SPDC signal spans from 842 nm to 878 nm with the conjugate idler photons spanning 7.32 μm to 11.38 μm, see Fig. 2. Figures 2a-c show the experimental results for different values of the displacement parameter: $\varepsilon_{exp}$=-1.0±0.1 mm, 0±0.07 mm, and 1.0±0.1 mm, respectively. Figures 2d-f show the corresponding theoretical patterns obtained with Eq. (10). For $\varepsilon$=0 mm, the spatial modes of SPDC photons generated from the first and second passes in the crystal fully overlap, see Fig. 2b. Therefore, no intensity modulation is observed. Once the crystal is slightly displaced, the interference fringes appear, see Figs. 2a,c – the effect is similar to the alignment of a conventional interferometer. The spacing between the interference fringes depends on the displacement lengths.

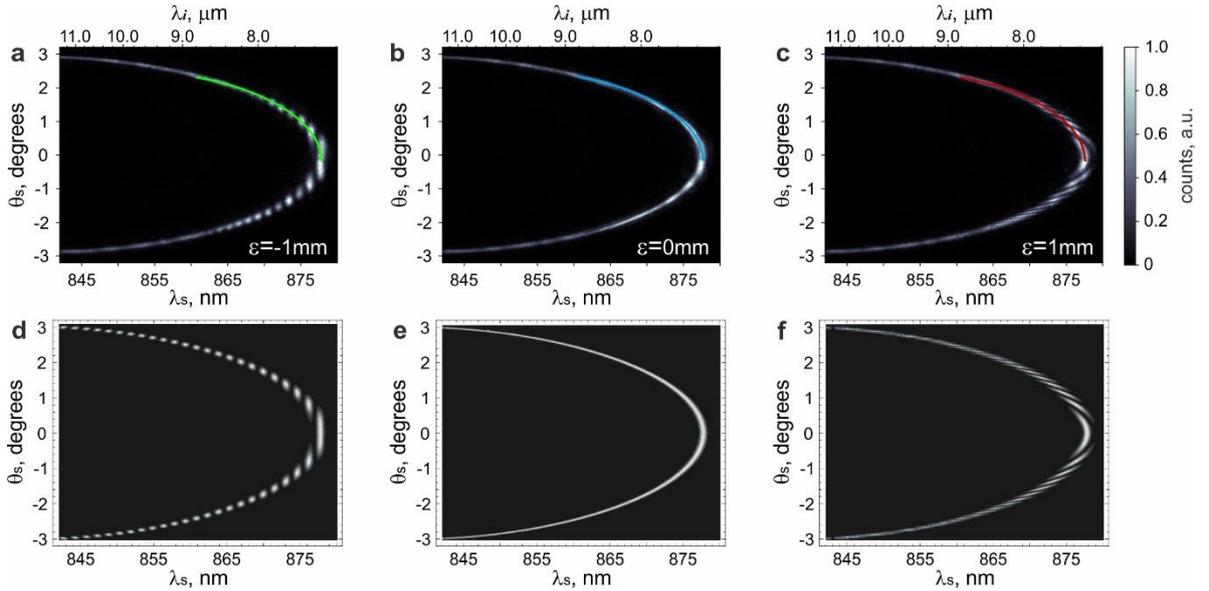

*Fig. 2 Interference pattern within the SPDC signal spectrum for different displacement values of the AGS crystal. a,d $\varepsilon$=-1 mm, b,e $\varepsilon$=0 mm and c,f $\varepsilon$=1 mm. Graphs a-c show the experimental data, and the graphs d-f show the theoretical calculations, which are in good agreement. Colored solid curves in a-c show the cross-section directions. Here and at later figures, the bottom and top x-axes show the detected and conjugate wavelengths, respectively.*

The interference fringes in Figs. 2a-c are observed within the external scattering angles of ±2.3 degrees. In contrast, theoretical simulations show the interference pattern across the whole range of observed angles. This is due to the limited numerical aperture of the OAPM, which cannot accommodate larger scattering angles $\theta_i$ of the idler photons. Neglecting the difference in the refractive indices at NIR and mid-IR ranges, the idler scattering angles are given by:

$$\theta_i = \theta_s \lambda_i/\lambda_s = \theta_s(\lambda_i/\lambda_p - 1), \qquad (13)$$

where $\lambda_{s,i}$ are the wavelengths of the signal and idler photons, respectively. For $\lambda_i$=8.8 μm and $\theta_s$=2.3°, see Fig. 2a-c, the quantum interference is only observable up to $\theta_i$=28.1° (0.49 rad). This agrees with the numerical aperture of our OAPM, which is given by $NA=d/(2f+\varepsilon)\approx 0.5$.

The interference pattern is experimentally observed in the wavelength range of signal photons of $\lambda_s$=865-878 nm. The corresponding range of idler photons is $\lambda_i$=8.37-7.32 μm. This range can be further extended by using OAPM with higher $NA$, or by using a longer pump wavelength[17], which will lead to a less non-degenerate SPDC regime and idler photons with smaller scattering angles, see Eq. (13).

We also note here that the interference fringes are exceptionally robust to mechanical and acoustic vibrations because all the three waves are co-propagating. Moreover, the interferometer does not need to be actively stabilized.

For further analysis of the interference pattern, we take cross-sections along the SPDC spectra, shown by solid lines in Fig. 2. The cross-sections are taken following the ellipse shape of the SPDC spectrum[40], where each wavelength $\lambda_s$ corresponds to a scattering angle:

$$\theta_s = \theta_{s0} \pm b\sqrt{1 - \frac{(\lambda_s - \lambda_{s0})^2}{a^2}}, \qquad (14)$$

where $\theta_{s0}$=0° and $\lambda_{s0}$=834 nm are the origins of the ellipse, $b$=3°, and $a$=44 nm are the minor and major axis lengths of the ellipse, respectively. We note the angular symmetry of the SPDC spectra. After noise subtraction and normalization of the cross-sections using the experimental data with $\varepsilon_{exp}$=0, the periodicity of the interference fringes agrees with the theoretical simulations, see Fig. 3. The best fit for the experimental data with $\varepsilon_{exp}$=-1.0±0.1 mm and $\varepsilon_{exp}$=1.0±0.1 mm correspond to theoretical values of $\varepsilon_{theory}$=-1.1 mm and $\varepsilon_{theory}$=-1.13 mm in Eq. (10), respectively.

As seen from Figs. 3a and 3b, the negative displacement *ε=-1 mm* yields interference fringes within a broader spectral range than for the case of *ε=+1 mm*. At the same time, the periodicity of the fringes is the same for the two cases. From Fig. 3b, it also follows that the visibility of the interference fringes decreases faster with the wavelength. This observation is consistent with the requirement for indistinguishability, given by Eq. (12).

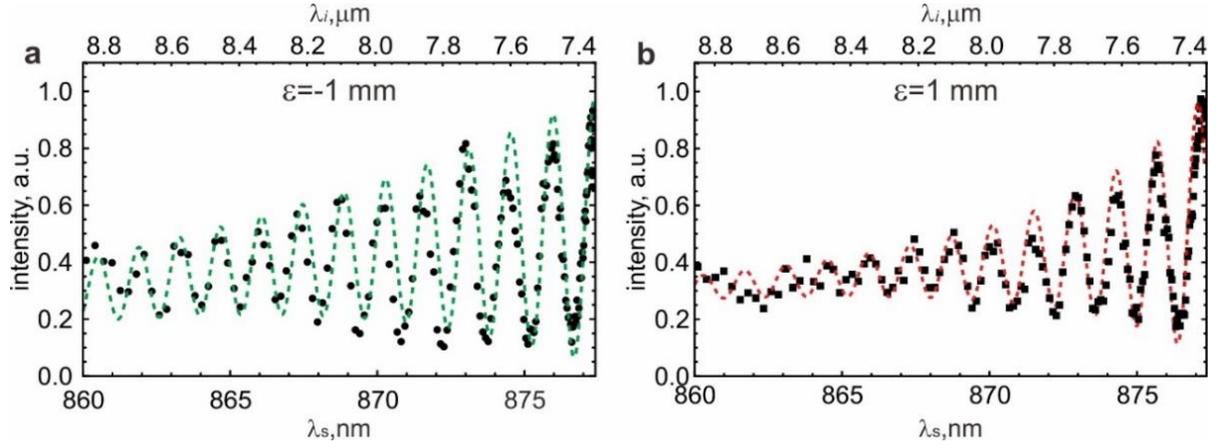

*Fig. 3 Cross-sections of the interference patterns in Fig. 2a,c. Cross-sections are taken along the upper half of the SPDC spectrum for the **a** ε=-1 mm, and **b** ε=1 mm displacement of the nonlinear crystal. Points correspond to the experimental data, and curves show fitting by the developed theoretical model.*

*Mid-IR gas spectroscopy*

Next, we study the absorption of nitrous oxide ($N_2O$) gas at around 7.8 μm. We set the crystal phase matching angle at 53.65°±0.02°, so that the SPDC signal spans from 841 nm to 878 nm with conjugate idler photons from 11.57 μm to 7.32 μm, see Fig. 4. Following our findings in the previous section, we set the displacement of the crystal in the negative direction, up to $\varepsilon_{exp}$=-1.8±0.1 mm, to obtain the fine interference fringes across a broad spectral range. In this case, the interaction length of the gas with the idler photons is $4f + \varepsilon$ = 201.4 mm.

Furthermore, we translate the center of the OAPM transversely to the pump beam. It enables us to detect clear interference fringes in the lower half of the spectrum. In this case, the interference fringes in the upper half of the SPDC spectrum fade at large angles. As a result, the interference pattern is observed at broader ranges with improved visibility at $\lambda_s$=860-878 nm and $\lambda_i$=8.87-7.32 μm for the signal and idler SPDC photons, respectively. In this case, the generated idler wavelengths cover the $N_2O$ absorption lines.

The reference interference pattern in Fig. 4a corresponds to results obtained with the evacuated chamber. Figures 4b,c correspond to results obtained with the $N_2O$ gas at a pressure of $P_1$=15±2 Torr and $P_2$=40±5 Torr, respectively.

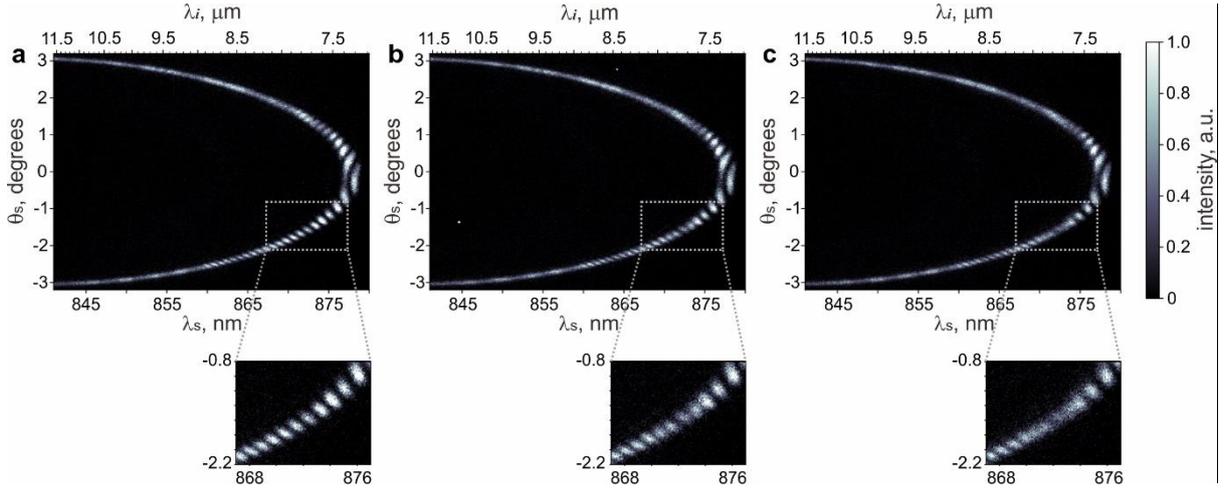

*Fig. 4 Interference patterns of SPDC signal photons. The chamber is filled with **a** vacuum; **b** 15 Torr $N_2O$ gas, and **c** 40 Torr $N_2O$ gas. Insets show the region of interest, where idler photons are absorbed by $N_2O$ at 7.7 μm and 7.9 μm. These regions show the modification of the interference pattern (reduction of the visibility and phase shift) due to gas absorption.*

Similar to the previous results, we analyze interference patterns with cross-sections taken according to Eq. (14). We detect changes in the visibility and phase of the interference fringes relative to the reference, see Fig. 5. These changes correspond to the resonant absorption and refraction by $N_2O$. Then, we calculate the change in visibility to infer the absorption coefficient according to Eq. (11), see Fig. 6a. Figure 6a shows excellent agreement of the measured absorption coefficient with the HITRAN database. The accuracy of the absorption coefficient measurements of our method is in the range 0.02-0.04 cm$^{-1}$.

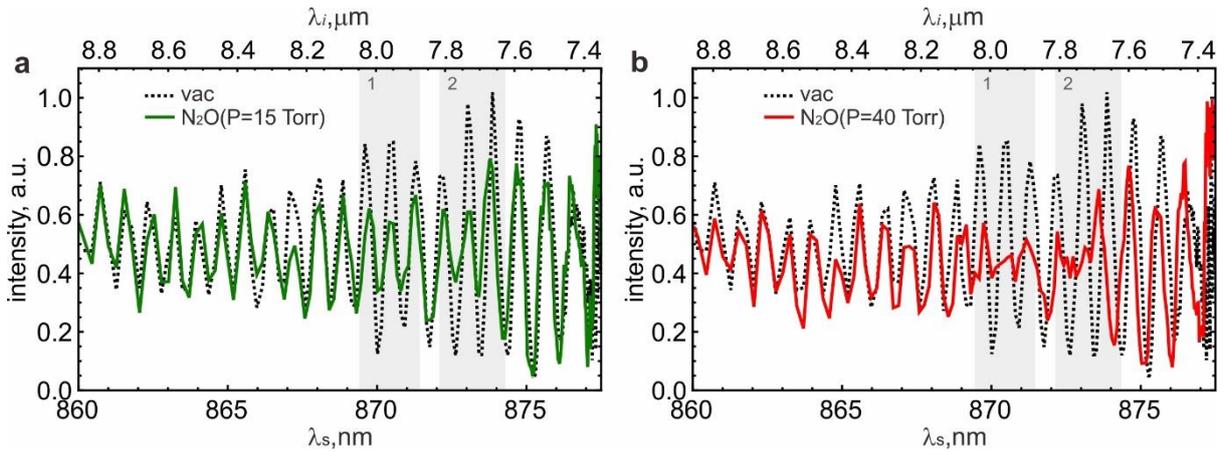

*Fig. 5 Cross-sections of the interference pattern in Fig. 4 taken along the lower half of the SPDC spectrum. The vacuum chamber is filled with **a** vacuum (black dashed line) or 15 Torr $N_2O$ (solid green line), and **b** vacuum (dashed black line) or 40 Torr $N_2O$ (solid red line). Highlighted gray regions ("1" and "2") show the regions of light absorption by the gas.*

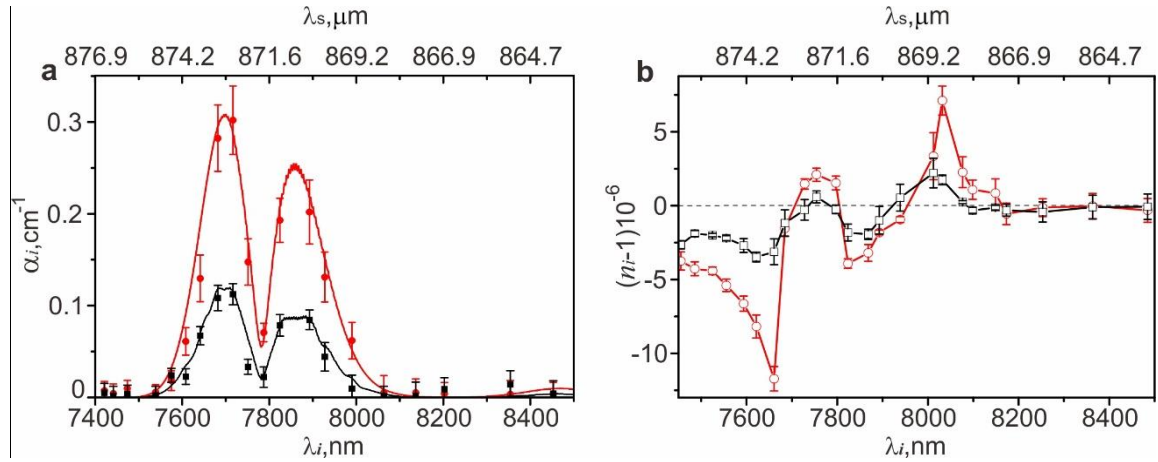

*Fig. 6 a Absorption coefficient and b refractive index of the $N_2O$ gas in the "fingerprint" mid-IR region. Points show experimental data, solid lines in a show HITRAN data for two values of pressure (red P=40 Torr, black P=15 Torr). Solid lines in b connect experimental data points for clarity.*

Next, assuming known values of the refractive indices of $N_2O$ in the NIR range[41], we retrieve its refractive index at 7.4-8.4 μm range. Figure 6b plots the change of the refractive index inside the vacuum chamber, which is measured with an accuracy of $1·10^{-6}$-$3·10^{-6}$.

The error bars in Figs. 6a,b are calculated taking into account the accuracy in determining values and positions of maxima/minima of interference fringes. Depending on the proximity of the wavelength of idler photons to absorption lines of the gas, the interference visibility fades that results in larger error bars.

The wavelength resolution of our system is ~50 nm (~8 cm$^{-1}$) in mid-IR range, which is ultimately limited by the spectral resolution of the visible/NIR range spectrometer. This result is on par with conventional FTIR techniques, which have a resolution at the order of 16 cm$^{-1}$ to 2 cm$^{-1}$. In our case, the achieved resolution is sufficient to distinguish the absorption doublet of $N_2O$ gas.

We now want to contrast the current work with earlier demonstrations[13,14]. In earlier works with two crystals, the interference fringes were observed only within ±1° of the SPDC spectrum with 250 nm IR bandwidth[13]. In contrast, the interference fringes in this work are observed across a much broader IR spectrum of 1.55 μm, see Fig. 4. It allows the analysis of gas absorption over a six times broader mid-IR range in a single shot. Note that with the current scheme, interference can be observed across the whole SPDC spectrum, provided the *NA* of the OAPM is large enough to collect all the scattered idler photons (see Fig. S1 in Supplementary Materials for the experiment with a lithium niobate crystal). Nonetheless, it is possible to cover a broader mid-IR range by changing the phase-matching angle of the crystal in the current system, see Fig. S2 in the Supplementary Materials.

Compared to previous works[13,14], we have also increased the interaction length of the gas with the idler photons. In earlier works, the interaction length was limited by the natural divergence of SPDC. In contrast, here, the interaction length is defined by the focal length of the OAPM and is increased by order of magnitude. Further increase of the interaction length is possible by using an OAPM with a longer focal distance or by implementing a multi-pass interferometer configuration.

Besides $N_2O$, many other molecules present absorption in the "fingerprint" mid-IR range, including sulfur dioxide ($SO_2$), Criegee intermediate $CH_2OO$[42], among many others. For example, $SO_2$ is a toxic gas, which needs to be monitored in industrial exhausts, and $CH_2OO$ is involved in global climate change research.

*Conclusion*

In conclusion, we demonstrated quantum interferometry with a versatile and stable interferometer configuration, where sample properties at "fingerprint" mid-IR range (7.32-8.87μm) can be revealed from the detection of the NIR light (860-878 nm). We performed theoretical derivations of the interference pattern and proof-of-concept mid-IR spectroscopy measurements of $N_2O$ gas using the proposed interferometer configuration. The method provides simultaneous measurement of the absorption coefficient and refractive index over a 1.55 μm mid-IR range, which is 6 times broader compared to previous works. The accuracy of measurements is 0.02 cm$^{-1}$ and $1\cdot10^{-6}$ for the absorption coefficient and refractive index, respectively, which are on par with existing commercial instruments.

This work extends the quantum interferometry technique's capability for fast and accurate probing of the properties of the media in the broad fingerprint region in mid-IR. Realization of the quantum interferometry in the "fingerprint" mid-IR range is a critical step towards the application of the method in real-world problems. We, therefore, believe that the demonstrated method has good potential for practical application both in the "fingerprint" mid-IR spectroscopy and imaging. While working on the manuscript, we became aware of the relevant work exploiting the FTIR principle for quantum spectroscopy in the mid-IR[43].

**Methods**

Our experimental setup is shown in Fig. S3 of Supplementary Materials. We use a continuous wave (cw) laser ($\lambda_P$=783.9 nm with FWHM=1.4 nm, 70 mW, Omicron) as the pump source for a silver thiogallate crystal ($AgGaS_2$ or AGS, *l*=2 mm, 51° cut angle, type-II phase matching, Altechna). We set the laser beam to have horizontal (H) polarization and the optical axis of the nonlinear crystal in the *xz* plane. The generated idler photons cover mid-IR wavelengths from 7 μm up to 10 μm, where most of the "fingerprint" absorption lines are located. In this case, the signal photon is generated at 850-882 nm.

After the crystal, the 90° OAPM (silver coated, $f$=50.8 mm, Edmund Optics) focuses $k$-vectors of the SPDC photons onto the surface of a flat mirror (silver coated, Thorlabs) with a diameter of 50.8 mm. After reflection from the flat mirror, the OAPM performs an inverse Fourier transformation, restoring the initial transverse distribution of the SPDC photons at the nonlinear crystal. As the OAPM does not introduce chromatic dispersion to the SPDC photons, dispersion compensating optics is not required. The flat mirror is carefully aligned at the focal distance from the OAPM. Displacement of the mirror from the focal distance decreases the visibility of the interference pattern. At the same time, the AGS crystal is mounted onto a translation stage to introduce shifts of $\varepsilon/2=\pm2.0$ mm in the lateral direction from the aligned position.

The reflected pump beam generates SPDC photons during a second pass through the crystal. The signal photons are focused into the slit of an imaging spectrometer (SpectraPro 2300i, Acton) by a three-lens system ($f_1$=100 mm, $f_2$=200 mm, $f_3$=125 mm). The first and second lenses are placed confocally, while the third lens is placed at a focal distance from the spectrograph slit. The wavelength-angular spectra are detected by an electron-multiplying (EM) CCD camera (iXon897, Andor, gain=295, 20s acquisition time, sensor temperature -70°C).

The interferometer is placed inside a vacuum chamber. The chamber is pumped down to a pressure of ~100 mTorr and then filled with a small amount of gas. The pressure inside the chamber is monitored by a pressure gauge (275 Mini-Convectron, Granville-Phillips). For our tests, we used nitrous oxide ($N_2O$) gas (purity 99.99%, National Oxygen), which has two distinctive absorption lines around 7.69 and 7.88 μm[31].

### Acknowledgments

The authors thank Dmitry Kalashnikov for discussions and advice on the experiment.

### Author contributions

A.P, D.T, and H.Y. contributed to the experimental work and building the setup. A.P performed the analysis of the results. A.P. and L.K. conceived the idea and designed the experiment. All authors contributed to the preparation of the manuscript.

### Funding

This work was supported by Agency for Science, Technology and Research (A*STAR) under the project #C21091702.

### Competing interests

The authors declare no competing interests.

# Supplementary materials for the manuscript:

# Broadband quantum spectroscopy at the fingerprint mid-infrared region

Anna V. Paterova[1,*], Zi S. D. Toa[1], Hongzhi Yang[1], and Leonid A. Krivitsky[1]

[1] Institute of Material Research and Engineering, Agency for Science Technology and Research (A*STAR), 138634 Singapore

*E-mail: Paterova_Anna@imre.a-star.edu.sg

**Experiment with a lithium niobate nonlinear crystal**

Using the proposed interferometer setup, we also performed an experiment using a lithium niobate (LNb) nonlinear crystal and 532 nm laser. In the experiment, the LNb crystal is oriented at $\theta=50°$ phase matching angle. In this case, the visible and mid-IR SPDC photons are generated at 580-618 nm and 3.8-5.5 μm, respectively. Signal after 5.5 μm starts to fade due to absorption by the crystal.

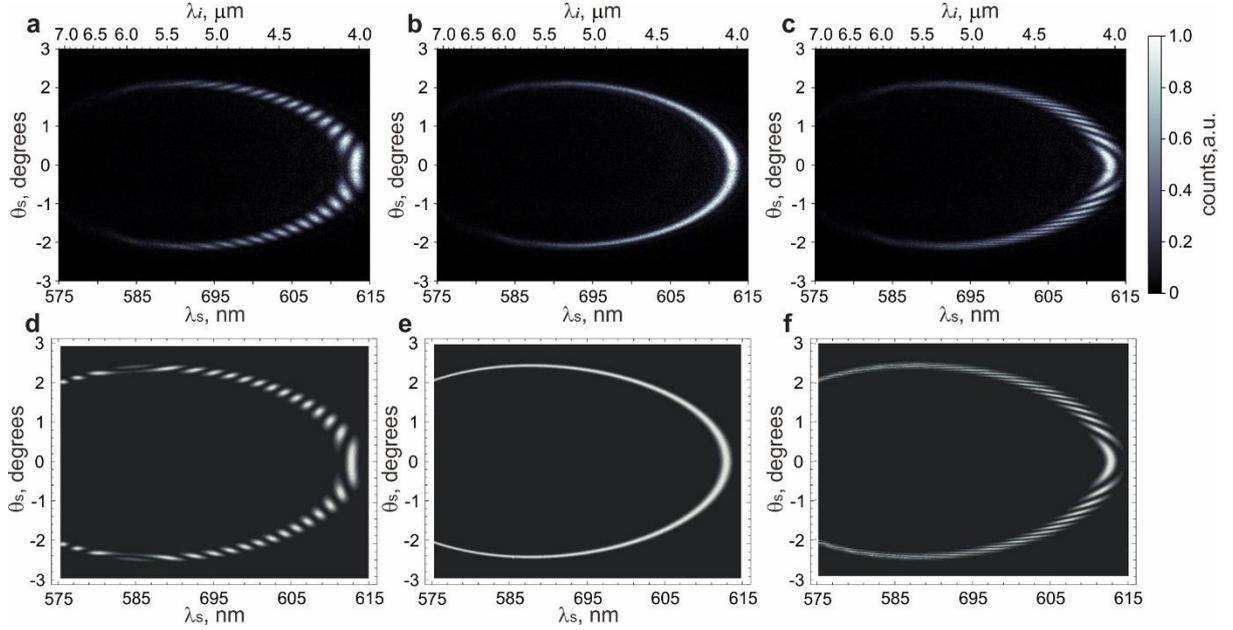

*Fig. S1 Interference pattern within the SPDC signal spectrum for different displacement positions of the LNb crystal: **a,d** $\varepsilon=-1$ mm, **b,e** $\varepsilon=0$ mm and **c,f** $\varepsilon=1$ mm. Graphs a-b represent the experimental data and the graphs e-f show the theoretical calculations.*

**Tuning the wavelength of the SPDC spectrum**

The range of interest in the mid-IR can be tuned via tilting the nonlinear crystal. Figure S2 shows the spectra for (a) 48.59°±0.02°, (b) 51.83°±0.02° and (c) 53.65°±0.02° orientation of the AGS crystal, where the collinear signal photons are detected at 860.1 nm, 871.3 nm and 877.34 nm, respectively. In this case, the interference fringes are observed up to 9.4 μm, see Fig. S2a.

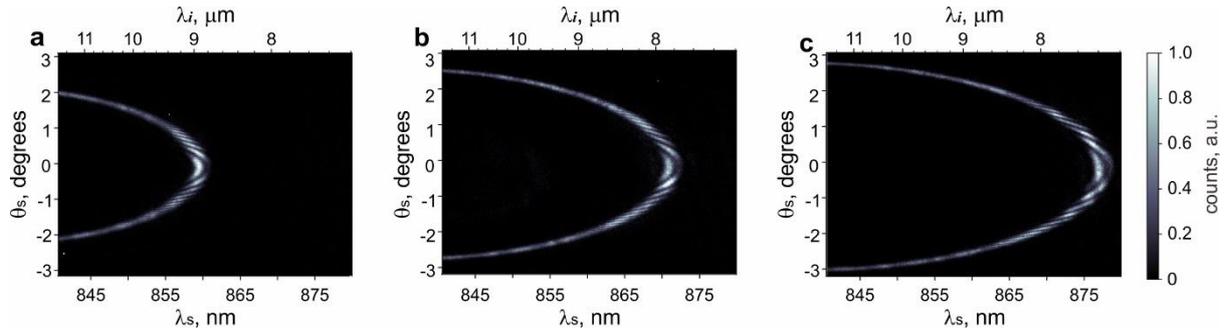

***Fig. S2*** *SPDC spectra from AGS nonlinear crystal at **a** 48.59°, **b** 51.83°, and **c** 53.65°±0.02° phase-matching angles.*

**Experimental setup**

Figure S3 shows the scheme of the experimental setup, which is discussed in detail in the "Methods" section of the manuscript.

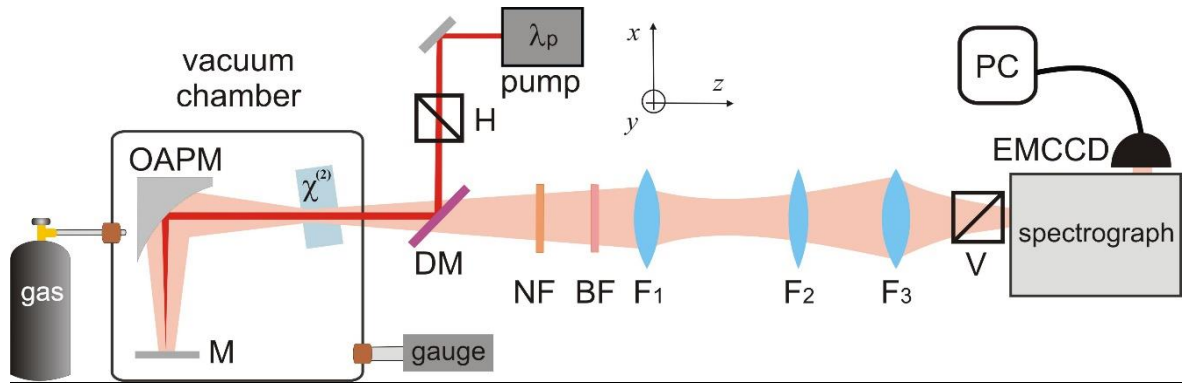

***Fig. S3*** *Experimental setup. A cw-laser is used to pump the AGS crystal, generating NIR and mid-IR photons via SPDC at the first pass of the pump. An off-axis parabolic mirror (OAPM) and flat (M) mirrors form the interferometer. Signal and idler photons are also generated by the reflected pump beam. The interference pattern of signal photons is captured in the wavelength-angular domain using a spectrograph and EMCCD camera preceded by a three-lens system ($F_1$, $F_2$, $F_3$). A notch (NF), bandpass (BF) filters and polarization filters (H, V) are used to filter the signal photons. The crystal and mirrors are placed inside the vacuum chamber purged by $N_2O$ gas. The gas concentration is measured by the pressure gauge.*